\documentclass[sigconf, nonacm, anonymous=false, screen, review=false, 10pt]{acmart}

\usepackage{xcolor}
\usepackage{pifont}
\usepackage{xurl}
\usepackage{booktabs}
\usepackage[ruled]{algorithm2e}
\usepackage{algpseudocode}
\usepackage{enumerate}
\usepackage{enumitem}
\usepackage[english]{babel}
\usepackage{blindtext}
\usepackage{subfig}
\usepackage{amsmath}

\usepackage{amssymb}
\usepackage{xspace}
\usepackage{multirow}
\usepackage{threeparttable}
\usepackage{soul}
\usepackage{float}
\usepackage{flafter}
\usepackage{comment}
\usepackage{graphicx}
\usepackage{caption}
\usepackage{balance}
\usepackage{textcomp}

\usepackage{xcolor}

\definecolor{SensingBlue}{HTML}{1479C9}
\definecolor{SensingBlueBG}{HTML}{EAF4FB}

\definecolor{EdgeGreen}{HTML}{6DAA3A}
\definecolor{EdgeGreenBG}{HTML}{EFF7E9}

\definecolor{EncoderPurple}{HTML}{7A3FA0}
\definecolor{EncoderPurpleBG}{HTML}{F3ECF8}

\definecolor{ShoreOrange}{HTML}{F28C28}
\definecolor{ShoreOrangeBG}{HTML}{FFF1E5}

\newcommand{\modulebox}[1]{%
  \raisebox{0.10ex}{\textcolor{#1}{\rule{0.62em}{0.62em}}}%
}

\DeclareRobustCommand{\sensingtag}{\modulebox{SensingBlue}}
\DeclareRobustCommand{\refreshtag}{\modulebox{EdgeGreen}}
\DeclareRobustCommand{\encodertag}{\modulebox{EncoderPurple}}
\DeclareRobustCommand{\shoretag}{\modulebox{ShoreOrange}}
\def\fig{Fig.\xspace}
\def\eqn{Eq.\xspace}
\def\tab{Tab.\xspace}

\def\ie{{\textit{i.e.}\xspace}} 
\def\eg{{\textit{e.g.}\xspace}}

\newcommand{\head}[1]{{\noindent \textbf{#1:}}}

\graphicspath{{./figures/}}
\graphicspath{{./pdf/}}
\DeclareGraphicsExtensions{.pdf,.jpeg,.png}

\ifodd 1
\newcommand{\com}[1]{\textbf{\color{red}(COMMENT: #1)}} 
\newcommand{\todo}[1]{\textbf{{\color{orange}(TODO: #1)}}}
\else

\newcommand{\com}[1]{}
\newcommand{\todo}[1]{}
\fi

\setcopyright{none}

\def\sysname{\textsc{CoraLive}\xspace}

\begin{document}

\title[\sysname: Ultra-Low-Power Underwater Vision for Long-term Coral Reef Monitoring]
{\sysname: Ultra-Low-Power Underwater Vision for Long-term Coral Reef Monitoring}

\begin{anonsuppress}

\author{Sitong Shen}
\affiliation{
  \institution{The University of Hong Kong}
  \country{Hong Kong SAR, China}
}
\email{shenst@connect.hku.hk}

\author{Chenshu Wu}
\affiliation{
  \institution{The University of Hong Kong}
  \country{Hong Kong SAR, China}
}
\email{chenshu@cs.hku.hk}
\end{anonsuppress}


\begin{abstract}
Effective coral reef protection and restoration demands continuous, long-term underwater monitoring. 
However, current solutions largely rely on fragmented, manual labor and costly infrastructure, or are severely limited by constrained battery lifespans. 
In this paper, we present \sysname, an ultra-low-power underwater camera system for long-term, continuous coral reef monitoring. 
\sysname presents a decoupled underwater-shore co-design that minimizes on-board computation and storage, offloading heavy reconstruction and segmentation tasks to offline shore-side processing. 
To achieve this, \sysname introduces a dual-rate sensing scheme combining frequent low-resolution frames with sparse high-fidelity samples and a content-adaptive update mechanism that extracts and stores only compact, task-specific data, eliminating the need to save raw video frames. 
We implement \sysname on a custom STM32U575-based dual-camera prototype with device-level computation and storage optimization. Compared with raw recording, \sysname reduces daily storage by $4.51\times$ and average power by $7.78\times$, extending battery lifetime from 20.4 to 159.5~days while retaining high-accuracy segmentation and reconstruction. Further validated in a real-world underwater deployment, we envision \sysname as a pioneering step for ubiquitous long-term coral reef underwater monitoring.

\end{abstract}

\maketitle

\section{Introduction}
\label{sec:intro}

\begin{figure}
\centering
\includegraphics[width=0.9\linewidth]{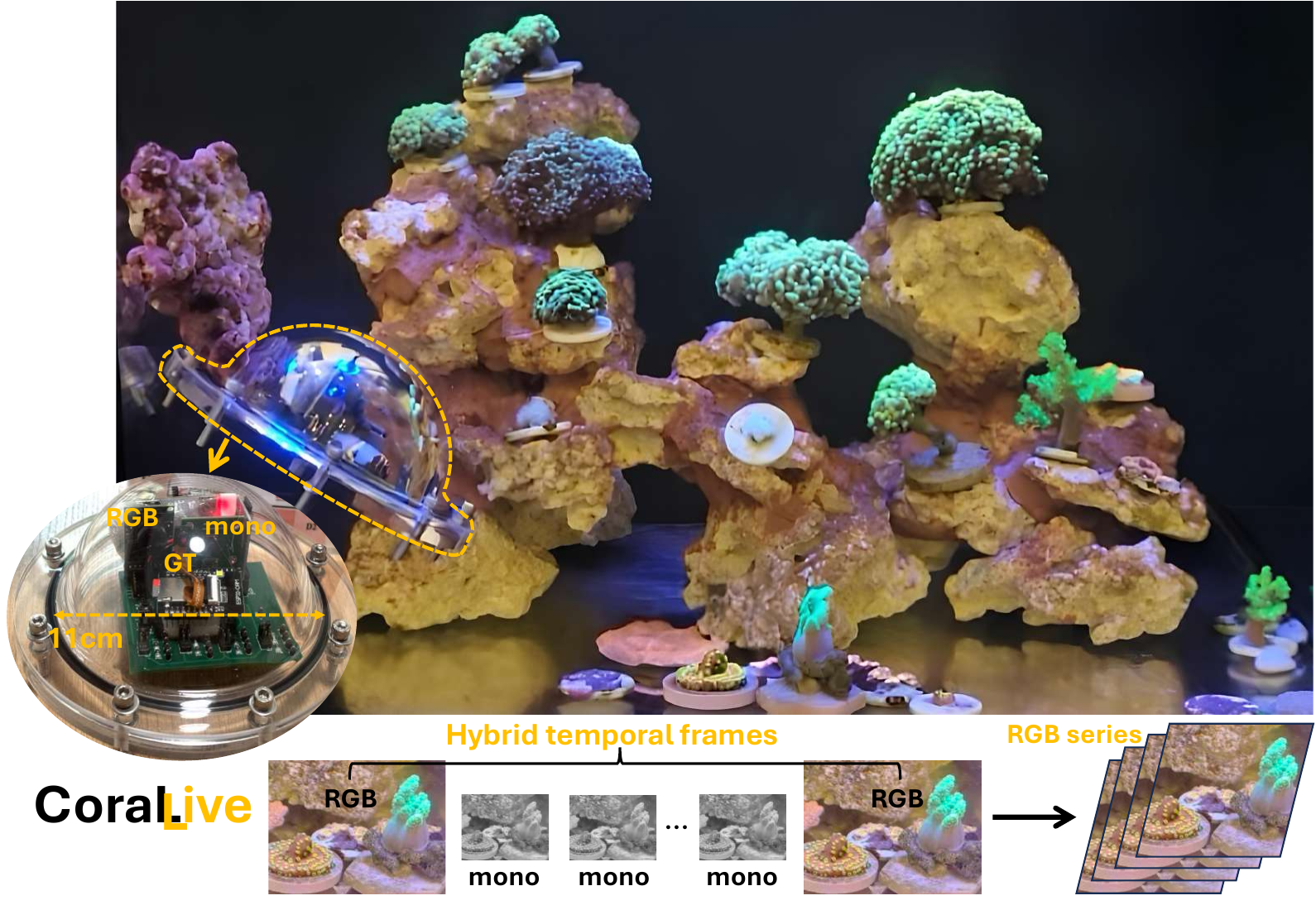}
\vspace{-1mm}
\caption{Overview of \sysname, a long-term, ultra-low-power underwater coral monitoring system.}
\label{fig:overview}
\vspace{-1mm}
\end{figure}

Coral reefs are essential ecological infrastructure that sustain marine biodiversity and provide broad ecological and socioeconomic benefits, including natural coastal protection~\cite{ferrario2014effectiveness}.
However, recurrent bleaching and large-scale mortality are accelerating reef degradation~\cite{hughes2018spatial,hughes2018global}, driving a global loss of approximately 14\% of coral between 2009 and 2018.
Active intervention can substantially reshape local reef cover~\cite{souter2021status,lange2024restoration}, 
yet ensuring the success of both protection and restoration strategies relies on \textit{long-term monitoring}~\cite{goergen2020coral,edmunds2024monitoring} because reef decline and recovery unfold over years to decades~\cite{hughes2018spatial,zeng2025inevitable,coralwatch}. 
Additionally, \textit{continuous} monitoring at a high frequency is equally critical to capture short-timescale biological processes. For example, coral spawning requires minute-scale tracking~\cite{takeuchi2022spawning}, while polyp motion and behavior occur at 10-second intervals~\cite{li2022spatiotemporal}, both necessitating frequent observations over the long term. 

Long-term continuous coral reef monitoring, however, remains challenging and costly. Existing reef monitoring spans three common approaches: (1) \textit{Diver surveys} offer detailed observations but are labor-intensive and thus infrequent, often conducted only annually or even at multi-year intervals~\cite{towle2022national}. They also require trained personnel and repeated underwater operations, increasing labor cost and safety risk~\cite{usgs2011diving}. (2) \textit{Satellite remote sensing} allows broad regional coverage but lacks underwater views with colony-level resolution to capture fine-grained changes~\cite{radford2024remote,hedley2016remote}. (3) \textit{Underwater platforms} can provide continuous in situ observations, but existing systems are expensive to build, operate, and scale. For example, NSF Ocean Observatories Initiative relies on $\sim$900~km of submarine power and telemetry cabling, while operating regional research vessels costs roughly \$29,000 per day~\cite{murphy2008ooi,ooiRegionalCabledArray,osuRCRVcost}. 
Recent advances in low-cost underwater cameras ~\cite{coralcitycamera,maurya2023augmented,naeem2024seascan, girdhar2023curee} offer a promising alternative for sustainable coral monitoring. However, their deployment duration and coverage are severely constrained by battery capacity and maintenance cost. 
We therefore ask the question: \emph{Can we build an ultra-low-power underwater vision system that sustains long-term, continuous coral monitoring with minimal intervention?}

To address this question, we separate underwater observation from immediate data access. 
Recognizing that these two tasks compete for a constrained power budget, our insight is that continuous data recording (making data available) should take priority over real-time communication (making data accessible). 
We therefore propose a decoupled edge-shore co-design, where observations are stored locally and efficiently, while intensive computation is offloaded to the shore side for offline processing after data retrieval. 
Built upon this architecture, we present \sysname, an ultra-low-power underwater vision system for long-term, continuous coral reef monitoring. 
As illustrated in \fig~\ref{fig:overview}, \sysname operates offline once deployed at an underwater site, compressing and storing data locally rather than streaming it. Once the system is retrieved, a sophisticated model is employed on the shore-side to recover high-fidelity visual views and perform complicated downstream tasks like reconstruction and segmentation. 
However, realizing such a system for ultra-long-term operation is non-trivial; \sysname features three distinct designs that address three major sources of power consumption to deliver an ultra-low-power system:

\head{Dual-Rate Visual Sensing}
Sampling the visual scenes at higher resolution and higher rates consumes more power. Reducing both directly extends deployment lifetime, but may sacrifice temporal coverage and miss coral dynamics. 
To save power consumption without impacting the quality of visual observations, \sysname introduces a \textit{dual-rate sensing} design by integrating two cameras: A high-resolution RGB camera captures high-fidelity visual scenes at a very low rate (\eg, every a few minutes), while a low-resolution monochrome camera samples the dynamics in between at a higher rate, \eg, every 10 s. 
The two streams complement each other, with the RGB frames capturing the slowly-varying color information while the monochrome frames supplementing local features like reef textures and structures, which together allow recovery of high-resolution, high-rate observations from the shore side with significantly lowered power consumption. 

\head{Content-Adaptive Partial-Frame Processing} 
The primary goal of coral reef monitoring is to preserve a continuous visual history over long-term deployments. 
However, directly saving every sampled full frame is power-hungry. 
In \sysname, we instead explore a \textit{content-adaptive refresh} mechanism to track incremental changes only rather than full frames, which remarkably reduces the amount of data to be processed and stored. 
Our insight is that, despite various temporal dynamics such as water motion, illumination changes, and moving organisms, underwater observations at a fixed view angle contain substantial reusable structure that persists over a considerably long period, allowing for frame compression. 
To achieve this, a core challenge couples with the low frame rate sensing,
where adjacent observations cannot simply be treated as nearly identical frames, creating a temporal structure distinct from typical video streams sampled at significantly higher rates, \eg, 24 fps. 
We present a lightweight encoder to maintain a compact, coral-aware visual state on-board. This reusable representation of the current scene will be updated only with newly observed information. Edge computation and storage therefore scale proportionally with scene change rather than full-frame size, while preserving the information required for later recovery.



\head{Device-Level Energy Optimization} 
Deployment lifetime depends on translating the above sensing and computational savings into lower device-level energy, which requires systematic integration and optimization. 
\sysname introduces a novel patch-wise state update policy to reduce the runtime cost of the neural execution for content-adaptive refresh.
Selective state updates skip encoder execution in stable regions and shrink the log produced by each observation. The compact records can be buffered in on-chip SRAM memory, allowing the microSD card to remain fully powered off between batched writes. 
This model-hardware co-design jointly reduces edge computation and hardware active time, substantially extending battery lifetime.


Besides the edge side, \ie, the underwater vision system, we also build a powerful shore-side model to decode dense visual data from the minimally stored information from the underwater counterpart. 
Our key idea is to preserve recoverable information underwater and recover dense visual outputs on shore.
After the underwater node is retrieved, the shore-side model causally integrates the sparse RGB anchors with the replayed sequence of compact updates to reconstruct the persistent visual state, which further supports downstream reconstruction and segmentation tasks.

We implement \sysname on a custom STM32U575-based dual-camera prototype with an HM01B0 monochrome camera, an OV2640 RGB camera, microSD storage, and built-in power measurement, as depicted in \fig~\ref{fig:overview}. 
Extensive experiments show that compared to baselines, \sysname reduces daily storage by $4.51\times$ and average power by $7.78\times$, extending battery lifetime by $7.82\times$, from 20.4 to 159.5~days using a 1000mAh battery, while retaining 0.825 segmentation mIoU and 34.192~dB RGB reconstruction PSNR. A seven-day underwater deployment further achieves 0.925 mIoU and 29.66~dB PSNR in coral scenes, consuming only 30.6~mAh. 
To our knowledge, \sysname is the first system to enable ultra-low-power, long-term, and continuous underwater coral reef monitoring, promising ubiquitous and scalable deployment. 

In summary, we make the following contributions:
\begin{itemize}[noitemsep, topsep=2pt, leftmargin=*]

\item We design a long-term coral monitoring system that preserves continuous reef observations while exploiting temporal and spatial redundancy to avoid repeatedly acquiring, processing, and storing complete visual information.

\item We develop an adaptive visual-state update method that carries useful information across observations and selectively refreshes changed or aging regions, enabling reliable visual history over long-term monitoring streams.

\item We customize a prototype that jointly optimizes sensing, computation, storage, and hardware activity. \sysname maintains strong coral monitoring performance while extending lifetime by around $8\times$.

\end{itemize}

The remainder of the paper is organized as follows. \S\ref{sec:opportunity} presents the design opportunities in fixed-view reef monitoring, and \S\ref{sec:design} describes the system design. \S\ref{sec:prototype} presents the hardware prototype and measurement setup, while \S\ref{sec:evaluation} evaluates task performance, system cost, and deployment lifetime. \S\ref{sec:discussion} discusses practical considerations and future extensions. \S\ref{sec:related} reviews related work, and \S\ref{sec:conclusion} concludes.

\begin{figure}[t]
    \centering
    \includegraphics[width=0.99\linewidth]{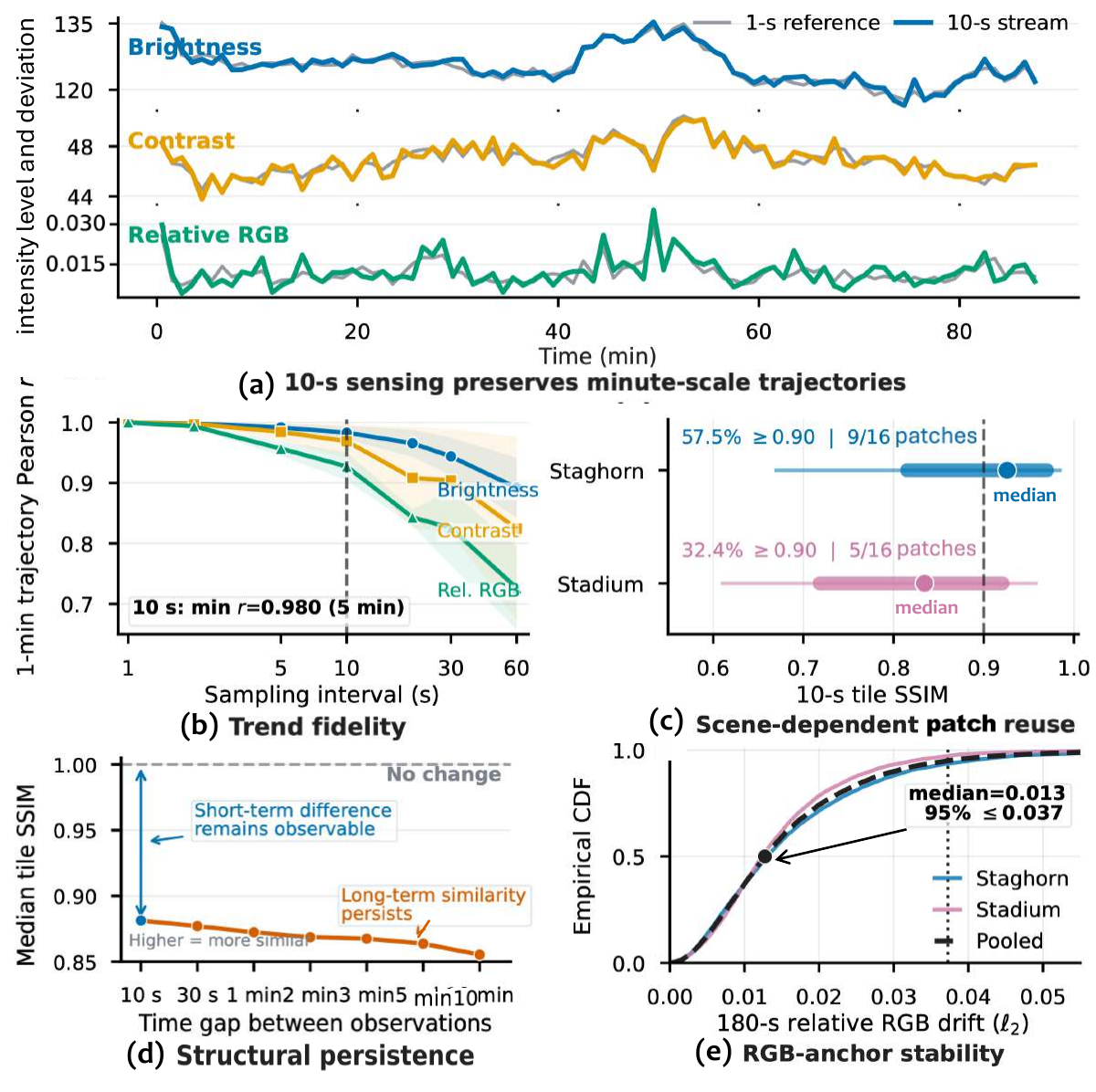}
    \vspace{-2.5mm}
    \caption{\textbf{Fixed-view reef video exhibits temporal, spatial, and color redundancy.} \mdseries (a)-(b) A 10-s stream preserves minute-scale trends. (c)-(d) Spatial similarity persists from seconds to minutes but varies by scene. (e) Relative color changes little over a 180-s RGB-anchor interval.}
    \label{fig:opportunity_keyidea}
    \vspace{-1.5mm}
\end{figure}

\section{Design Opportunities}
\label{sec:opportunity}

Before designing the underwater node, we first ask what visual information long-term coral monitoring actually needs to preserve. Image-based coral monitoring relies on two important visual cues. First, coral structure, including colony extent, boundaries, and morphology, allows researchers to estimate coral cover and colony area and to track growth, mortality, and recovery over time~\cite{edmunds2024monitoring,khen2024inter}. Second, color provides complementary information about coral condition. For example, CoralWatch, a global coral-monitoring program used in around 80 countries, uses a standardized Coral Health Chart to compare observed coral colors with reference colors for assessing changes in coral health~\cite{coralwatch,pedraza2026beyond}. Thus, a coral reef monitoring system needs to preserve both \emph{structure}, capturing coral extent, boundaries, and morphology, and \emph{appearance}, capturing coral color changes over time. 

We next examine how these visual contexts evolve given a fixed-view observation point. Successive observations often contain substantial redundancy, while illumination, water motion, and moving organisms introduce short-lived visual variation. We therefore characterize three questions: \emph{(1) how frequently the reef must be observed, (2) how much spatial structure information can be reused across observations, and (3) how often the color appearance must be refreshed}. To answer these questions, we analyze two fixed-view Coral City Camera recordings~\cite{coralcitycamera}, Staghorn Rubble Zone and Stadium, totaling 3.9~h, which helps us identify potential opportunities for reducing redundant sensing, computation, and storage. 

\head{Lower-rate sensing preserves minute-scale trends}
The native recordings are sampled at ${\sim}30$ FPS, producing a dense temporal stream of about 1,800 frames per minute. For an ultra-low-power underwater node with limited on-board compute and memory resources, continuously acquiring and processing this volume of data would impose substantial computation and data-movement overhead. We therefore ask whether the sensing cadence can be reduced while still preserving the temporal information needed for monitoring. 

We start by comparing a 1-FPS stream directly against the native recordings to verify that it closely preserves their minute-scale scene trends. We quantify trend similarity using the Pearson correlation coefficient $r$~\cite{benesty2009pearson}, which quantifies the degree of linear correspondence between two temporal trajectories, with $r=1$ indicating perfect positive correlation. The 1-FPS stream achieves $r>0.99$ for brightness, contrast, and relative-RGB trends at the 1-min scale, and $r>0.997$ at the 5-min scale. 
Using 1 FPS as the starting point, we gradually reduce the sampling rate and compare the performance in \fig~\ref{fig:opportunity_keyidea} (a)-(b). As seen, denser sampling preserves more temporal information but also increases the sensing and processing burden. At a 10-s interval, brightness, contrast, and relative-RGB trajectories remain strongly correlated with the reference, with 1-min correlations of $0.983$, $0.969$, and $0.927$, respectively, using only one-tenth as many observations; and with 5-min minimum remains $0.980$. This preliminary result reveals substantial temporal redundancy: 10-s sampling retains minute-scale scene trends at a substantially lower sensing rate. We therefore adopt 10~s as the fast sensing cadence, balancing temporal coverage against the computation and energy constraints of the underwater node.

\head{Spatial structure remains reusable across observations} Using a 10-s sampling interval, the system still produces 8,640 observations per day, making full-frame processing and storage costly for long-term deployment. We therefore ask whether every observation must be treated as a complete new frame. To analyze this, we divide each frame into a $4\times4$ grid and compare corresponding reef regions using the Structural Similarity Index Measure (SSIM)~\cite{wang2004image}. Using $0.90$ as a strong reference, a median of 9/16 regions in the Staghorn dataset and 5/16 in the Stadium dataset remain structurally similar after 10~s; over longer intervals, the median SSIM decreases only from $0.881$ at 10~s to $0.855$ at 10~min, as illustrated in \fig~\ref{fig:opportunity_keyidea}(c)-(d). This analysis reveals two key properties: the fraction of structurally similar regions varies across scenes, yet this similarity persists over multiple observations.

\head{Color needs less frequent sampling}
We next examine whether color must be refreshed at the same rate as the structure. We normalize each frame to the same global grayscale intensity, preserving only changes in the relative composition of the R/G/B channels and compare each observation with its latest RGB anchor. At a 180-s anchor interval, the median relative-RGB drift is $0.013$, with $95\%$ of observations below $0.037$, as shown in \fig~\ref{fig:opportunity_keyidea}(e). Relative color changes negligibly over the evaluated 180-s interval, supporting less frequent RGB refresh than the fast structural stream. 
This observation motivates our dual-rate design: 
While RGB recording is too costly and monochrome sensing alone cannot preserve absolute color information, we can combine the two with periodic RGB frames as anchors and monochrome samples in between for continuous tracking. 
We use 10s interval for the low-power monochrome camera while 180~s for the RGB channel, which is a conservative refresh interval rather than a claimed optimum.

\head{Design implication}
In summary, these measurements reveal four key observations for long-term fixed-view reef monitoring. (1) Minute-scale scene trends remain observable with 10-s sampling, revealing substantial temporal redundancy. (2) Many corresponding regions remain structurally similar across observations, but the fraction varies across scenes. (3) This spatial similarity persists over multi-minute intervals but gradually decreases, allowing retained information to remain useful while requiring periodic refresh. (4) Relative color changes little over 180~s, allowing RGB to be refreshed less frequently than structural observations. These observations motivate our system design, as detailed next.

\vspace{-2.5mm}
\section{System Design}
\label{sec:design}

\sysname is a long-term, ultra-low-power dual-camera coral monitoring system that combines frequent monochrome sensing with periodic RGB observations and reuses unchanged visual information over time. It preserves recoverable information underwater and recovers dense visual outputs on shore. The underwater node maintains a spatially aligned visual state, updates only regions containing new information, and logs the resulting state updates rather than complete images. After retrieval, the shore replays these states to recover the visual-state history, and combines it with the RGB anchors to reconstruct frequent high-resolution RGB observations and coral-monitoring outputs. This design reduces underwater computation and energy consumption while retaining the visual information required after deployment.
Guided by the temporal, spatial, and color redundancy characterized in \S\ref{sec:opportunity}, the monochrome camera observes reef structure every 10~s, while the RGB camera provides color information every 180~s. A lightweight microcontroller performs selective on-board processing and local storage so that the resulting reductions in computation and data volume translate into lower device-level energy.

\begin{figure*}[t]
    \centering
    \includegraphics[width=0.92\textwidth]{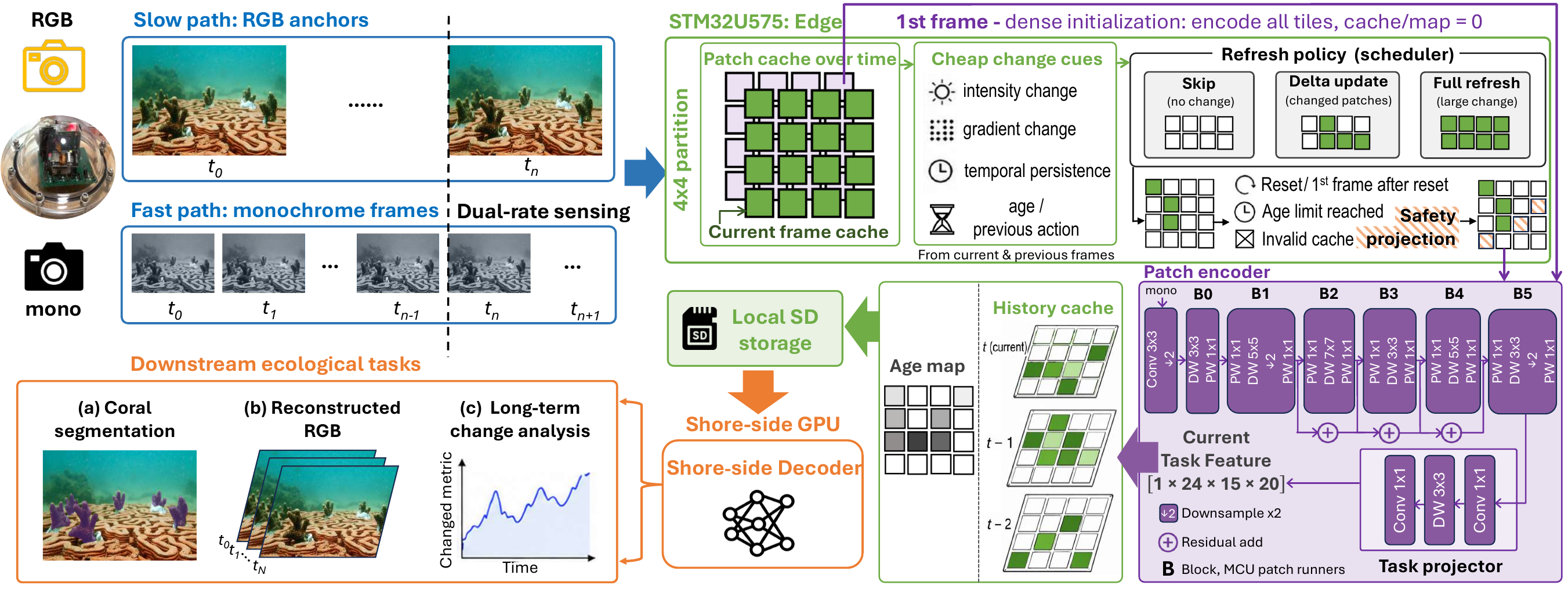}
    \vspace{-2mm}
    \caption{\textbf{Dual-rate persistent-state edge-shore architecture.} \mdseries
    \sensingtag~ Dual-rate image acquisition.
    \refreshtag~On-board change cues, refresh control, state history, freshness safety, and local logging.
    \encodertag~Patchwise MCU encoding and task-state projection.
    \shoretag~GPU-side decoding and ecological outputs.
    The first observation or a reset initializes the complete cache; later observations refresh only selected patches and reuse the remaining state.}
    \label{fig:arch}
    \vspace{-0.5mm}
\end{figure*}

Long-duration underwater cameras typically retain observations locally powered by batteries, because it is cheap and reliable to preserve data~\cite{greene2020coralcam,bicknell2016camera,coralcitycamera}. We therefore choose a local microSD card for underwater storage.
Although modern microSD cards provide ample capacity, writing large visual payloads remains costly on MCU-class platforms. Practical throughput is limited by bus bandwidth and buffering, preventing the MCU from sustaining the card's advertised peak rate~\cite{broell2023iot}. Larger payloads consequently keep the storage subsystem active for longer. Our design therefore exploits fixed-view redundancy \emph{before data reaches storage}: frequent monochrome observations track current structure, sparse RGB anchors preserve color, and only selected regions of a persistent visual state are computed and logged, offloading dense computing to the shore side after data retrieval.

\subsection{Design Overview}
\label{sec:design_overview}

Figure~\ref{fig:arch} organizes the system into four coordinated modules. 
\sensingtag~The sensing module acquires frequent monochrome observations and sparse RGB anchors. 
\refreshtag~The refresh module extracts low-cost change cues, and determines which regions require an update. 
\encodertag~The encoding module executes the selected encoder patches and converts the resulting feature map into compact states. 
\shoretag~The shore-side module receives the states and RGB anchors logged to the microSD and manually retrieved from the deployment, and replays them through decoding models for coral-related analysis.

The underwater node maintains visual representations across successive monochrome observations rather than processing every frame independently. 
Each observation is divided into fixed spatial regions, or \emph{patches}, so that different parts of the scene can be updated independently. The first observation initializes the complete representation using a lightweight on-device encoder. 
For each later observation, a lightweight \emph{scheduler} computes change cues across observations to determine which patches require new visual information and which can reuse their previously stored representations. 
Selected patches by our scheduler are refreshed by the encoder, while the remaining regions are carried forward from the previous observation. 

These selective changes are stored as compact updates rather than complete frames and buffered locally for microSD storage.
Periodic RGB anchors are stored alongside these updates to preserve color and appearance information unavailable from the monochrome stream. 
After the underwater node is retrieved, the shore replays the logged data in temporal order to reconstruct the persistent visual-state sequence and combines it with the RGB anchors to reconstruct dense RGB observations and coral-analysis outputs.

Together, the edge-shore architecture provides the persistent state used when the next monochrome observation arrives, and reconstructs the high-resolution RGB observations over time, allowing \sysname to sustain ultra-low-power coral monitoring. The edge-side change cues calculation, scheduler refresh decision, selective patch execution, and shore-side recovery are detailed in the following sections.

\subsection{Dual-rate Sensing and Persistent State}
\label{sec:persistent_state}
The submerged node uses two complementary sensing paths: a fast path uses a low-power monochrome camera to track rapidly changing intensity and scene dynamics, while a slow path sparsely activates the higher-power RGB camera to capture slowly varying color information as anchors.

\head{Fast path} We use a HM01B0 camera captures a $160\times120$ monochrome observation every 10~s. This stream provides current evidence on reef boundaries, texture, local contrast, and scene layout without repeatedly activating the higher-resolution, high-power color path. 

\head{Slow path} The OV2640 camera captures a $640\times480$ RGB JPEG anchor every 180~s, as monochrome sensing alone cannot preserve absolute chromatic state. Underwater attenuation and scatter are wavelength-dependent, and distinct RGB appearances may therefore produce similar grayscale observations~\cite{akkaynak2018revised,akkaynak2019sea}. Each anchor preserves a recent high-resolution appearance and color reference at substantially higher per-frame energy cost.

\head{Persistent visual state} Frequent monochrome observations remain redundant because much of a fixed reef scene changes little over 10~s. Rather than processing every observation independently, the node retains a compact visual representation across time: regions that change are recomputed, while stable regions reuse their previous representations. We call this retained representation the \emph{persistent visual state}. The next subsection explains how these regions are defined, selected for update, and carried forward over time.

\subsection{Content-Adaptive Refresh}
\label{sec:refresh_safety}

Our fast sensing path does not process the entire monochrome image at every observation. Instead, we use a lightweight on-board encoder that converts image regions into compact visual features, while the node retains these features across observations as its persistent visual state. Each observation is divided into a fixed $4\times4$ grid of 16 spatial patches, so that each patch has a corresponding stored feature. The first observation is processed in full to initialize these features. For each later observation, the system determines which stored features can still represent the current scene and which need to be recomputed.

This decision is made before running the fast encoder. Using the monochrome observation and recent history for each patch, a lightweight \emph{scheduler} produces a tentative $4\times4$ refresh mask indicating which patches should be recomputed. Mandatory freshness rules may add patches whose stored features have become too old or unreliable. The resulting mask controls execution: selected patches are recomputed and replace their stored features, while the remaining patches reuse the previous ones. 
Each observation proceeds through lightweight change-cue extraction, refresh decision, freshness enforcement, selective fast-encoder execution, and state update. We detail these stages below.

\head{Low-cost change cues} Before running the neural encoder, the scheduler computes eight lightweight features for each patch in the $4\times4$ grid. These features summarize monochrome intensity change, gradient change, accumulated residual change, patch age, the previous refresh decision, and coarse temporal context. All features are causal and available before neural inference.

\head{Content-adaptive refresh} For each of the 16 patches, the scheduler uses these low-cost cues to judge whether the patch has changed enough to require a new neural feature. Patches showing clear or persistent change are refreshed, while stable patches reuse the stored feature in the cache. The 16 refresh-or-reuse decisions form a $4\times4$ raw refresh mask $\widetilde{M}_t$. The scheduler decides which patches to refresh using only the current monochrome observation and lightweight temporal history, before running the neural encoder.

\head{Mandatory safety freshness}
Adaptive refresh alone may repeatedly defer updates when frame-to-frame changes remain small. Over many observations, however, these small differences can accumulate, causing the retained feature to drift from the current scene and gradually reducing its reliability.
We therefore force a refresh in three cases: the patch age reaches six observations, the cache is initialized after a reset, or the cached feature becomes invalid because accumulated change exceeds an empirical threshold.
The final refresh mask is $M_t=\widetilde{M}_t\lor M_t^{\mathrm{age}}\lor M_t^{\mathrm{reset}}\lor M_t^{\mathrm{invalid}}$, where $\lor$ denotes logical OR. A patch is therefore recomputed whenever the scheduler or any safety condition requests it.
\vspace{-2.5mm}
\subsection{Patchwise Update and Local Logging}
\label{subsec:patch_execution}

The previous subsection determines which image patches require updates. We now describe how this $4\times4$ refresh mask drives MCU execution and how the resulting updates are stored for later shore-side processing. A selected patch is recomputed and replaces its previous representation, whereas an unselected patch performs no corresponding neural computation and retains its stored representation.

\head{Patchwise encoder execution}
To execute selected regions at the edge, we partition the fast encoder into four callable stages as in \fig\ref{fig:arch}: B2-B5, named by the backbone block at which each stage ends. B2 contains the input convolution and blocks 0-2, while B3-B5 contain the subsequent blocks. Each stage output is spatially cached, $16\times30\times40$ for B2-B4 and $24\times15\times20$ for B5, so later observations can update selected regions without recomputing the full feature maps. Initialization or reset processes the complete image through B2-B5; for other observations, only patches selected by $M_t$ are recomputed, while the remaining regions retain their cached values.

\head{Context-preserving cache update} A selected patch cannot be processed using only its strict image boundary because convolutional features near that boundary also depend on neighboring content. For each selected patch, the MCU therefore extracts a slightly larger region that includes the required surrounding context, called a \emph{halo}. The corresponding stage processes this enlarged region, but only the central region belonging to the selected patch is written back to the stored feature map. Regions extending beyond the image boundary are zero-padded. This crop-execute-paste procedure preserves the convolutional context required by each stage while preventing one patch update from overwriting the state of neighboring patches.

\head{Fast task state}
After selective stage updates, the assembled B5 stage feature passes through a lightweight $1\times1$--depthwise-$3\times3$--$1\times1$ projector that produces the current $24\times15\times20$ fast task state, while a separate $2\times15\times20$ state preserves appearance information for shore-side reconstruction. Four patch-level status maps: age, stale status, refresh activity, and reset status, are retained as metadata $\mathrm{Meta}_t$ to guide shore-side recovery.

\head{Compact local log}
Selective computation alone cannot reduce storage energy if every 10-s observation still triggers a microSD write. 
We keep the persistent state in MCU SRAM and buffer compact updates before activating the microSD for batched storage.
A raw $160\times120$ monochrome frame occupies 19.2~KB, whereas a fast-stream update is approximately 1.6~KB and becomes smaller when fewer patches are refreshed. This buffering is important because repeatedly power-cycling the card at the 10-s sensing cadence is inefficient: microSD reinitialization can take roughly 40--150~ms while drawing about 110~mA~\cite{bradley2020optimising}. Accumulating multiple observations before each SD activation therefore reduces both startup and write activity; we quantify the resulting energy savings in Section~\ref{sec:evaluation}.
Each logged record contains the refreshed feature regions together with the refresh mask, state metadata $\mathrm{Meta}_t$, observation index, and latest RGB-anchor reference. 
After retrieval, the shore replays updates, replacing refreshed regions while carrying reused regions forward to reconstruct the persistent visual-state sequence for RGB reconstruction and coral analysis.

\vspace{-2.5mm}
\subsection{Shore-Side Ecological Recovery}
\label{sec:shore_recovery}
As mentioned above, our design prioritizes continuous underwater data collection over real-time data communication. Consequently,  \sysname excludes the communication module and relies on manual retrieval (\eg, by divers) after long-term operations. Given \sysname's drastically extended battery life, this constraint is practically negligible as physical retrievals are required only annually or even less frequently. 

After retrieval, the shore replays the logged patch updates in chronological order to recover the same task-state history maintained underwater. At frame $t$, this replay provides a $24\times15\times20$ fast task state $F_t$, a $2\times15\times20$ reconstruction-side state $S_t$, and the refresh-status maps $\mathrm{Meta}_t$. At a high level, these inputs have complementary roles: the task state provides current reef structure, the sparse RGB anchors provide color and texture, the monochrome history describes temporal change, and the refresh metadata indicates how recent each region is. The shore combines this information to produce a coral mask and a high-resolution RGB frame.

\begin{figure}[t]
    \centering
    \includegraphics[width=\linewidth]{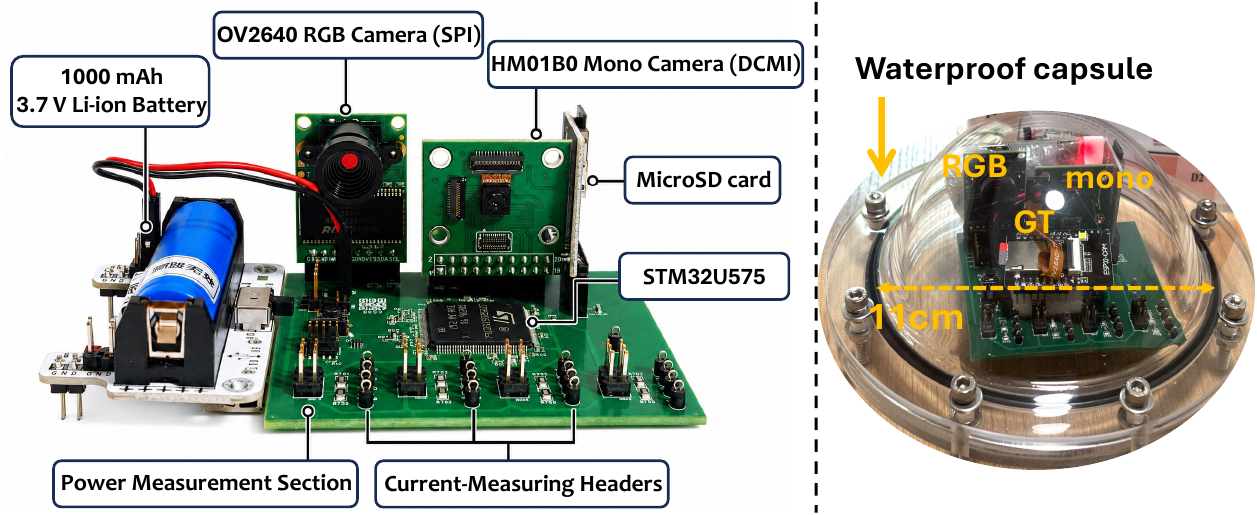}
    \caption{\textbf{Left:} \mdseries Custom \sysname hardware platform with power measurement section.
    \textbf{Right:} \mdseries Entire system with waterproof capsule and ground truth camera.}
    \label{fig:hardware}
\end{figure}

\head{RGB-guided coral segmentation} Pixel-level coral masks localize coral extent and colony boundaries, enabling coral-cover estimation, morphology analysis, and temporally consistent tracking~\cite{zheng2024coralscop,zheng2025coralvos,wong2025coralscop}. We first decode $F_t$ into a structural segmentation estimate using an Atrous Spatial Pyramid Pooling (ASPP)-style context block~\cite{chen2017deeplab}. Four parallel convolutions with dilation rates $1,2,4,6$ capture reef structures at different spatial scales, followed by refinement blocks with channel widths $96\!\rightarrow\!96\!\rightarrow\!64\!\rightarrow\!32$ and a final $1\times1$ convolution. The output is two unnormalized class scores, or logits, for coral and background, denoted by $L_t^{\mathrm{base}}$.

In parallel, the latest RGB anchor available at or before frame $t$ is processed by a pre-trained MCUNetV2-style RGB encoder and a lightweight projector to produce an appearance feature $A_t$ that preserves color and texture cues unavailable from the monochrome stream alone. A shore-side semantic correction head $\mathcal{C}_{\mathrm{seg}}$ combines $A_t$, the fast feature $F_t$, the task-state-guided temporal feature $H_t^{\mathrm{guided}}$ defined below, and refresh-status metadata $\mathrm{Meta}_t$:
\begin{equation}
L_t
=
L_t^{\mathrm{base}}
+
\mathcal{C}_{\mathrm{seg}}
\!\left(
H_t^{\mathrm{guided}},F_t,A_t,\mathrm{Meta}_t
\right),
\label{eq:rgb_guided_segmentation}
\end{equation}
where $L_t$ denotes the final coral/background logits and $\mathrm{Meta}_t$ contains the patch age, stale, active, and reset maps.

\head{Task-state-guided RGB reconstruction} High-resolution RGB observations preserve reef texture, local contrast, colony morphology, and chromatic appearance needed for human inspection and downstream health/change analysis~\cite{towle2022national,radford2024remote,kopecky2023quantifying}. We reconstruct these observations using a causal NeuriCam-style temporal propagation backbone~\cite{veluri2023neuricam}. Here, \emph{causal} means that reconstruction at frame $t$ uses only observations and RGB anchors available at or before $t$, never future frames. The backbone produces a 64-channel propagated visual feature $H_t^{\mathrm{prop}}$ and a base RGB reconstruction $\hat{I}_t^{\mathrm{base}}$.

To incorporate the current MCU state, the $24$-channel fast feature $F_t$ and a small $2$-channel reconstruction-side feature $S_t$ are mapped to a $64$-channel task feature $T_t$, matching the width of $H_t^{\mathrm{prop}}$. Because different patches may have been refreshed at different times, a freshness gate also uses $\mathrm{Meta}_t$ to control how strongly the task feature modifies each region:
\begin{equation}
T_t=\mathrm{Adapter}([F_t,S_t]),
\label{eq:decoder_task_injection}
\end{equation}
\begin{equation}
H_t^{\mathrm{guided}}
=
H_t^{\mathrm{prop}}
+
P_0\!\left(
\mathcal{G}(H_t^{\mathrm{prop}},T_t,\mathrm{Meta}_t)\odot T_t
\right).
\label{eq:decoder_task_prop}
\end{equation}
Here, $[\cdot]$ denotes channel-wise concatenation, $\mathrm{Adapter}$ maps the combined task state to 64 channels, $\mathcal{G}$ is a sigmoid freshness gate with outputs in $[0,1]$, $\odot$ denotes element-wise multiplication, and $P_0$ is a zero-initialized projection. The resulting $H_t^{\mathrm{guided}}$ therefore combines the propagated visual history with the current task state while accounting for how recently each patch was refreshed.

For final RGB refinement, $H_t^{\mathrm{guided}}$, $F_t$, $S_t$, $\mathrm{Meta}_t$, and the current one-channel monochrome observation are concatenated, giving $64+24+2+4+1=95$ input channels. Two 128-channel convolutional blocks and progressive upsampling produce a three-channel residual $R_t$. The final RGB reconstruction is
\begin{equation}
\hat{I}^{\mathrm{RGB}}_t
=
\hat{I}^{\mathrm{base}}_t
+
0.1\tanh(R_t),
\label{eq:rgb_residual_refinement}
\end{equation}
where $\hat{I}^{\mathrm{RGB}}_t$ is the reconstructed RGB frame and $\tanh(\cdot)$ bounds each residual value to $[-1,1]$. The factor $0.1$ limits the correction magnitude in the normalized RGB space, preserving appearance propagated from the RGB anchors while allowing the current task state to restore local changes.


\section{Prototype Design}
\label{sec:prototype}

\begin{figure*}[t]
    \centering
    \includegraphics[width=0.92\linewidth]{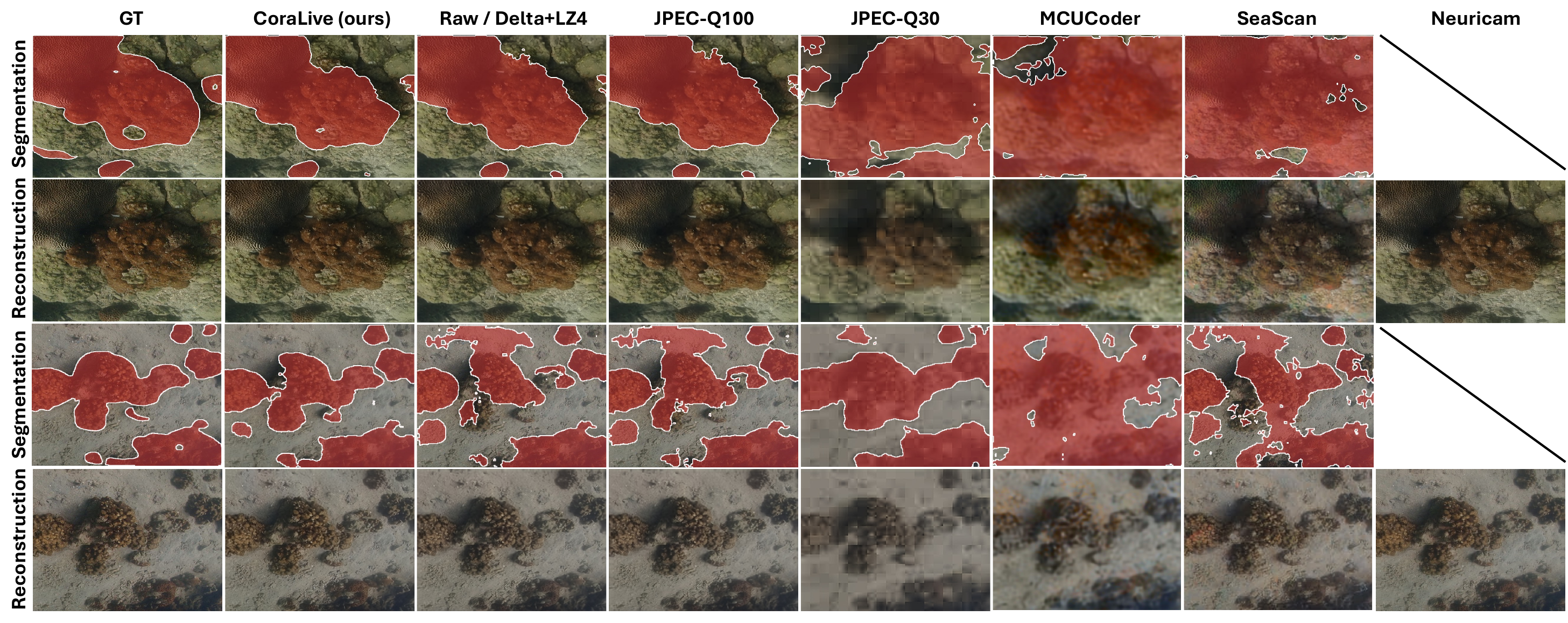}
    \vspace{-1mm}
    \caption{\textbf{Qualitative comparison of segmentation and RGB reconstruction.} \mdseries Representative results from \sysname and baselines on two reef scenes. Red overlays show coral segmentation.}
    \label{fig:comparison}
\end{figure*}

\head{Fast monochrome path} \fig~\ref{fig:hardware} (left) shows our dual-camera sensing platform, where a single STM32U575 controls both cameras and local storage at 160~MHz with its internal SMPS mode enabled. The fast path uses a Himax HM01B0 to capture $160\times120$ grayscale images every 10~s, with TIM1 providing the camera clock and I\textsuperscript{2}C handling sensor configuration. During capture, DCMI receives the pixel stream and GPDMA transfers it directly into SRAM without per-pixel CPU handling. Because the sensor uses a 4-bit DCMI bus, each 8-bit pixel requires two transfers, producing 38.4~KB of interface data that is packed in place into a 19.2-KB RAW8 image. This DMA-based path leaves most of each 10-s interval available for computation or low-power operation.

\head{Sparse RGB anchor path} The slow sensing path uses an OV2640-based ArduCAM module to capture $640\times480$ color images. The module performs JPEG compression in hardware and keeps the encoded image in its on-board FIFO buffer. For each RGB capture, the MCU clears the FIFO, starts the camera, waits for the capture-complete signal, and reads the JPEG payload over SPI in 4-KB chunks. The FIFO is important for the memory-constrained MCU because the complete compressed image remains on the camera module during capture; the MCU only needs a small transfer buffer while reading it out. We acquire one RGB anchor every 3~min, providing periodic color information without continuously operating the higher-cost color sensing path.
\begin{table}[t]
\centering
\footnotesize
\setlength{\tabcolsep}{2.5pt}
\caption{Key current and runtime parameters.}
\label{tab:power_runtime}
\begin{tabular}{lccc}
\toprule
\textbf{Component} & \textbf{Operating state} & \textbf{Current} & \textbf{Duration} \\
\midrule
STM32U575
  & Active
  & $6.7$ mA
  & always-on \\
STM32U575
  & SRAM retention
  & $4.3~\mu$A
  & between events \\
HM01B0
  & QQVGA capture
  & $0.24$ mA
  & $33.3$ ms / 10 s \\
OV2640
  & VGA JPEG capture
  & $70$ mA
  & $70.7$ ms / 180 s \\
microSD
  & Sleep
  & $1.2$ mA
  & between writes \\
microSD
  & Write (with initialization)
  & $50$ mA
  & $46$ ms / $19.2$ KB \\
\bottomrule
\end{tabular}
\end{table}

\head{Local storage and SD power gating}
Both sensing paths share a 32-GB microSD card connected to the STM32U575 through a 4-bit SDMMC interface. The fast path stores compact edge representations, while the slow path stores RGB JPEG anchors. Rather than keeping the card continuously powered, the node buffers data in SRAM and activates the microSD only when a write is required. After the buffered data are safely written, the card is powered off again. In our evaluated configuration, compact logging extends the interval between SD activations to approximately 29~min, allowing the card to remain power-gated for most of the deployment. This substantially reduces storage energy because the microSD draws considerably more power when active than the sensing and compute components.

\begin{figure*}[t]
    \centering
    \includegraphics[width=\linewidth]{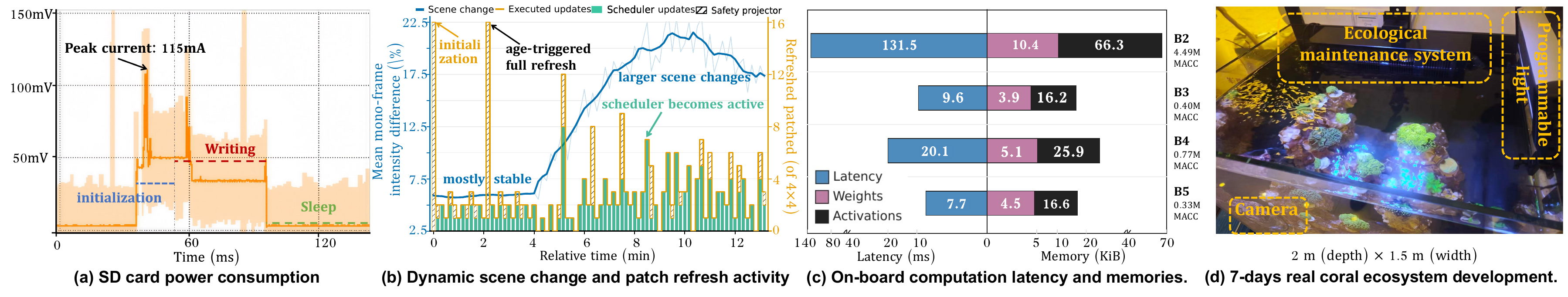}
    \caption{System measurements and real-world deployment.}
    \label{fig:computing}
\end{figure*}

\head{Integration and synchronization} A single STM32U575 controls both sensing paths, avoiding the clock synchronization required by multi-controller camera systems. At each 3-min RGB anchor, the MCU starts the OV2640 capture and arms an HM01B0 snapshot from the same firmware-controlled timing sequence. The paired observations are assigned the same logical timestamp, providing an explicit association between the fast monochrome observation and its RGB anchor. Our custom PCB integrates the MCU, camera interfaces, microSD storage, power distribution, and dedicated current-measurement headers on one board. Separate measurement points for the MCU, SD card, monochrome camera, and RGB camera allow us to measure the energy cost of each subsystem without modifying the sensing data path. The complete platform is powered by a rechargeable 3.7-V, 1000-mAh Li-ion battery, which provides a fixed energy budget for our end-to-end lifetime evaluation. The entire system is housed in a hemispherical waterproof enclosure, as shown in \fig~\ref{fig:hardware} (right). We use ESP32-CAM module with OV2640 RGB camera, which has $640\times480$ resolution, as ground truth camera. It is powered by another power rail, which has the same battery model as our system. We use GPIO in ESP32-CAM connected to our STM32U575 MCU to synchronize the data capture process.

\section{Evaluation}
\label{sec:evaluation}

We evaluate the proposed system in terms of algorithm performance, coral-monitoring utility, storage and energy efficiency, and battery lifetime. We compare against representative compression and underwater-imaging baselines and conclude with a seven-day end-to-end deployment in a real seawater environment.

\subsection{Experimental Setup}
\label{subsec:setup}

\head{Data authorities} CoralMask~\cite{zheng2024coralscop} provides 34,289/3,809/830 train/validation/test image--mask pairs for coral-domain pre-training, using $640\times480$ RGB and corresponding $160\times120$ monochrome inputs. 
CoralVOS~\cite{zheng2025coralvos} provides temporal data, split into two views, resized from $1280\times720$ to $640\times480$, and converted to $160\times120$ monochrome observations; we use 64,426 samples for edge training, 12,004 for shore-decoder training and validation, and 18,516 for final evaluation. 
For fixed-view adaptation, we use 42,000 Staghorn/Stadium observations~\cite{coralcitycamera}, including 5,000 human-finalized masks split by recording segment for training/development/evaluation and 37,000 CoralSCOP-LAT pseudo-labels~\cite{wong2025coralscop} used only for fine-tuning.
The seven-day deployment is reserved for end-to-end validation.

\head{Encoder training} Both visual encoders are initialized from the ImageNet-pretrained \texttt{mcunet-in2} backbone~\cite{lin2021mcunetv2}; the monochrome first convolution is initialized by averaging its RGB input kernels. CoralMask pre-training uses AdamW for 40 epochs with cross-entropy plus $0.5$ Dice loss. The monochrome encoder uses a learning rate of $3\times10^{-4}$ and batch size 24, while the RGB encoder uses $8\times10^{-5}$ and batch size 8. Temporal task-state training on CoralVOS then uses AdamW with a base learning rate of $6\times10^{-5}$, batch size 24, and six epochs, with additional update-ratio and refresh-mask losses weighted by 0.02 and 0.04, respectively.

\head{Refresh scheduler training} We train the refresh scheduler using self-critical REINFORCE with Adam, learning rate $10^{-5}$, rollout length 4, and logit scale 2.5. The objective is $\mathcal{L}_{\mathrm{sched}}=-A\log\pi+0.05D_{\mathrm{KL}}-0.001\mathcal{H}$, where $A=R_{\mathrm{sampled}}-R_{\mathrm{greedy}}$ and the reward balances task quality against executed patch updates. Patch-age, reset, and invalid-cache refreshes are deterministic safety actions; the maximum patch age is six.

\head{Shore-side training} The shore models are trained using the refresh-and-reuse states generated by the edge encoder and scheduler. The final RGB reconstruction model uses 70,244 training samples and 6,186 samples for evaluation. Final refinement uses AdamW with $lr=5\times10^{-5}$ for the task-state modules and $5\times10^{-6}$ for the reconstruction tail, with batch size 4 for two additional epochs. The reconstruction loss is $\mathcal{L}_{\mathrm{rec}}=\mathcal{L}_{\mathrm{MSE}}+0.15\mathcal{L}_{\mathrm{Charb}}+0.03\mathcal{L}_{\nabla,1}+0.03\mathcal{L}_{\mathrm{SSIM}}+10^{-4}\mathcal{L}_{\mathrm{res}}$.

\begin{table}[t]
\centering
\footnotesize
\setlength{\tabcolsep}{5.0pt}
\renewcommand{\arraystretch}{1.12}
\caption{System cost breakdown across design stages.}
\label{tab:system_cost_breakdown}
\begin{tabular}{lccc}
\toprule
\textbf{Configuration}
& \textbf{Data I/O}
& \textbf{Edge compute}
& \textbf{Avg. power} \\
& \textbf{(MB/day)}
& \textbf{(ms/obs.)}
& \textbf{(mW)} \\
\midrule
Dense RGB
& 530.84
& 0
& 10.05 \\

Dual-rate RAW
& 195.38
& 0
& 5.29 \\

+ Task encoder
& 97.21
& 168
& 5.17 \\

+ Adaptive refresh
& 43.32
& 50
& 4.63 \\

+ SD power gating
& \textbf{43.32}
& \textbackslash
& \textbf{0.68} \\
\bottomrule
\end{tabular}
\end{table}

\head{Baselines} We implement six baselines for comparison: 
\textit{RAW} stores every $160\times120$ monochrome observation and $640\times480$ RGB anchors without compression. 
\textit{Delta+LZ4} first encodes the frame difference and then applies lossless LZ4 compression, a lightweight method well suited to MCU-class devices~\cite{collet2011lz4}.
\textit{JPEC-Q100} and \textit{JPEC-Q30} use JPEC, a minimal grayscale JPEG encoder written in C for lightweight embedded use, at two quality settings ($Q=30$ and $Q=100$). 
\textit{MCUCoder} is a learned image-compression model for resource-constrained IoT devices; we use its official pretrained INT8 model, originally evaluated on $224\times224$ images~\cite{hojjat2025mcucoder}. 
We also reproduce \textit{NeuriCam} is a state-of-the-art dual-camera system that combines frequent low-power monochrome sensing with sparse high-resolution RGB frames for color reconstruction~\cite{veluri2023neuricam}. 
\textit{SeaScan} is a state-of-the-art underwater imaging system combines backscatter communication and SeaNet-based reconstruction~\cite{naeem2024seascan}. For a consistent comparison, RAW, Delta+LZ4, JPEC, and MCUCoder use the same 10-s monochrome and 180-s RGB sensing schedule as \sysname; NeuriCam and SeaScan retain their published sensing protocols.
\begin{table*}[t]
\centering
\footnotesize
\setlength{\tabcolsep}{3.0pt}
\renewcommand{\arraystretch}{1.10}

\caption{End-to-end comparison of system cost and coral-monitoring quality across baselines.}
\label{tab:system_baseline_comparison}

\begin{tabular}{ccc cccc cccc cc}
\toprule

\multirow[c]{3}{*}{\textbf{System}}
& \multirow[c]{3}{*}{\shortstack{\textbf{Original}\\\textbf{resolution}}}
& \multirow[c]{3}{*}{\shortstack{\textbf{Frame}\\\textbf{rate}}}
& \multicolumn{4}{c}{\textbf{System Cost}}
& \multicolumn{4}{c}{\textbf{Segmentation}}
& \multicolumn{2}{c}{\textbf{Reconstruction}}
\\

\cmidrule(lr){4-7}
\cmidrule(lr){8-11}
\cmidrule(lr){12-13}

&
&
&
\shortstack{\textbf{Bytes/day}\\\textbf{(MB)}}
& \shortstack{\textbf{Energy/day}\\\textbf{(mWh)}}
& \shortstack{\textbf{Avg. power}\\\textbf{(mW)}}
& \shortstack{\textbf{Battery life}\\\textbf{(days)}}
& \shortstack{\textbf{mIoU}\\\textbf{$\uparrow$}}
& \shortstack{\textbf{FG IoU}\\\textbf{$\uparrow$}}
& \shortstack{\textbf{Recall}\\\textbf{$\uparrow$}}
& \shortstack{\textbf{Precision}\\\textbf{$\uparrow$}}
& \shortstack{\textbf{RGB PSNR}\\\textbf{(dB)}}
& \shortstack{\textbf{SSIM}\\\textbf{$\uparrow$}}
\\

\midrule

RAW data
& 640x480
& 10 s
& 195.4
& 127.0
& 5.29
& 20.4
& 0.851
& 0.762
& 0.853
& 0.876
& 32.41
& 0.974
\\

Delta+LZ4
& 640x480
& 10 s
& 182.1
& 125.9
& 5.246
& 20.6
& 0.851
& 0.762
& 0.853
& 0.876
& 32.41
& 0.974
\\

JPEC-Q100
& 640x480
& 10 s
& 152.5
& 127.8
& 5.32
& 20.3
& 0.811
& 0.760
& 0.859
& 0.869
& 32.116
& 0.972
\\

JPEC-Q30
& 640x480
& 10 s
& 42.0
& 15.9
& 0.66
& 163.4
& 0.514
& 0.479
& 0.824
& 0.534
& 27.29
& 0.909
\\

Neuricam
& 640x480
& 1/15 s
& 504 GB
& > 3960
& > 165
& < 0.65
& -
& -
& -
& -
& 30.83
& 0.962
\\

SeaScan
& 320x240
& 187 s
& 9.2
& 15.1
& 0.63
& 171.4
& 0.322
& 0.249
& 0.902
& 0.256
& 28.23
& 0.952
\\

MCUCoder
& 224×224
& 10 s
& 60.6
& 179.1
& 7.46
& 14.5
& 0.455
& 0.464
& 0.908
& 0.486
& 25.34
& 0.554
\\

\textbf{\sysname}
& \textbf{640x480}
& \textbf{10 s}
& \textbf{43.3}
& \textbf{16.2}
& \textbf{0.68}
& \textbf{159.5}
& \textbf{0.825}
& \textbf{0.767}
& \textbf{0.847}
& \textbf{0.885}
& \textbf{34.192}
& \textbf{0.981}
\\

\bottomrule
\end{tabular}

\end{table*}
\head{Evaluation protocol} We evaluate all methods on the same datasets and under the same downstream task protocol. We report mIoU, foreground IoU, recall, and precision for segmentation; PSNR and SSIM for RGB reconstruction; and stored or transmitted bytes/day, measured energy/day, average power, and battery lifetime for system cost. RGB-anchor frames are excluded from reconstruction metrics to avoid inflating PSNR and SSIM. All MCU-side neural models use INT8 inference (8-bit integer arithmetic) for efficient on-device execution, while FP32 (32-bit floating-point arithmetic) is reported only as an accuracy reference. RAW, Delta+LZ4, and JPEG use their native implementations, while MCUCoder, SeaScan and NeuriCam follow their published pipelines.

\subsection{System Overall Performance}
\label{subsec:overall}

\head{Battery lifetime calculation}
Following prior battery models~\cite{bradley2020optimising,kerasiotis2010battery}, we compute the time-averaged load current as $I_{\mathrm{load}}=(I_{\mathrm{sleep}}t_{\mathrm{sleep}}+I_{\mathrm{active}}t_{\mathrm{active}})/T$. The lifetime is
\begin{equation}
L=\frac{\alpha C_{\mathrm{battery}}}{I_{\mathrm{load}}},
\label{eq:battery_lifetime}
\end{equation}
where $C_{\mathrm{battery}}$ is the nominal battery capacity. We use the common empirical factor $\alpha=0.7$~\cite{bradley2020optimising,farnellBatteryLife}. 
We use a $3.7$-V, $1000$-mAh LiPo battery for lifetime evaluation. According to \eqn~\eqref{eq:battery_lifetime}, and adopting the factor $\alpha=0.7$, the effective battery capacity used in the lifetime calculation is $700$~mAh. We directly measure both the current draw and the duration of the major units' active and sleep states from differential shunt-voltage traces captured using an Analog Discovery oscilloscope~\cite{digilentAD2}, with firmware runtimes additionally cross-checked in STM32CubeIDE. The microSD draws approximately $1.2$~mA in sleep and predominantly $\sim50$~mA during writes, with brief peaks of $100$-$110$~mA, as shown in \fig~\ref{fig:computing}(a). A $19.2$-KB raw mono (1 frame) write takes approximately $46$~ms. The remaining parameters are summarized in \tab~\ref{tab:power_runtime}.

\head{Storage reduction and power savings} \tab~\ref{tab:system_cost_breakdown} breaks down the on-board storage and energy costs across the major system stages. Dual-rate sensing reduces persistent I/O by 63.2\%, while task-state encoding and adaptive refresh further reduce it to 43.32~MB/day. The payload therefore decreases from 530.84 to 43.32~MB/day, allowing observations to accumulate in SRAM and extending the SD write interval to approximately 29~min. Rather than repeatedly reinitializing and waking the microSD, which incurs substantial overhead as discussed in~\S\ref{subsec:patch_execution}, the system keeps the card fully powered off between batched writes. SD power gating therefore reduces average power by $7.78\times$ relative to dual-rate RAW logging. \sysname consumes 16.2~mWh/day, corresponding to a projected battery lifetime of 159.5~days.

\head{Adaptive refresh behavior}
\fig~\ref{fig:computing}(b) shows how the refresh policy responds over time. During stable periods, few optional regions are refreshed; as visual change increases, the scheduler activates additional regions, while age-based safety rules periodically force refreshes. Across the evaluated sequences, the policy refreshes 2.35 of 16 regions per observation on average, corresponding to 85.3\% reuse. Optional refresh activity increases with visual change ($\rho=0.532$).

\begin{figure*}[t]
    \centering
    \includegraphics[width=0.92\linewidth]{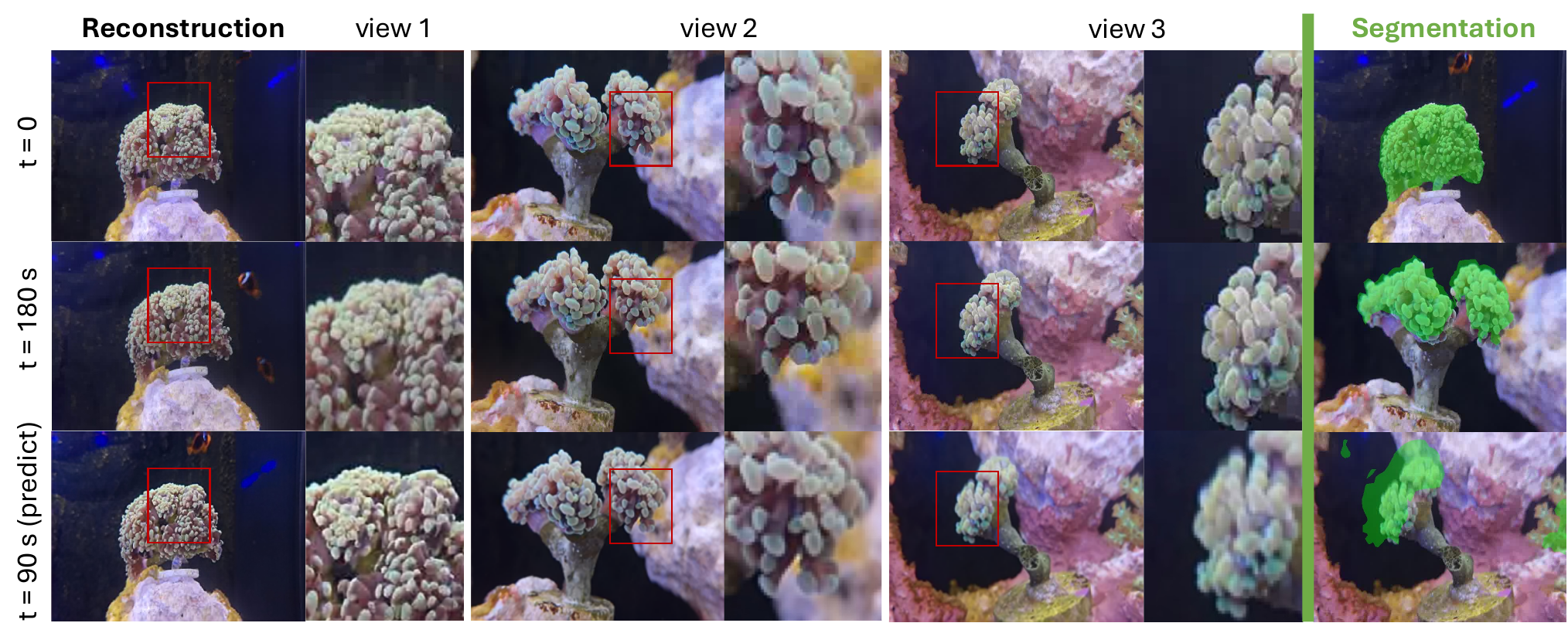}
    \vspace{-2mm}
    \caption{\textbf{Seven-day deployment results.} \mdseries Representative RGB reconstruction and coral segmentation from three fixed views. Intermediate frames are reconstructed from the sparse RGB anchors and fast task state; red boxes highlight local details.}
    \label{fig:case}
    \vspace{-2mm}
\end{figure*}

\head{On-board computation cost}
\fig~\ref{fig:computing}(c) reports the measured latency and activation footprint of the INT8 B2--B5 stages on the STM32U575. A complete sparse B2--B5 refresh requires 5.98M MACC and 168.8~ms, with B2 accounting for 75.0\% of the MACCs and 77.9\% of the measured runtime. B2 also has the largest activation footprint at 66.3~KiB, showing that avoiding unnecessary early-stage execution provides the largest opportunity for on-board compute savings.

\head{INT8 quantization and deployment fidelity}
We apply static post-training quantization (PTQ)~\cite{banner2019post,nagel2021white} using MinMax calibration~\cite{onnxruntime_quantization}, which determines activation ranges from the observed extrema of calibration data. Activations use asymmetric UINT8 quantization with scale and zero point, while weights use per-channel INT8 quantization. Using the same downstream model, the FP32 encoder achieves 0.837 mIoU, 0.740 foreground IoU, 0.864 recall, 0.865 Dice, 39.753~dB PSNR, and 0.990 SSIM. After INT8 quantization, the largest absolute change among segmentation metrics is 0.027; mIoU decreases by 0.012 and recall by 0.017. SSIM decreases by only 0.009 to 0.981, while PSNR decreases by 5.561~dB to 34.192~dB. These results support the full-INT8 configuration used for MCU deployment.

\head{System cost} The customized sensing node costs approximately \$30.9, and a waterproof enclosure costs less than \$20, keeping the total deployed system cost below \$51.

\subsection{End-to-End System Comparison}
\label{subsec:system_comparison}

\tab~\ref{tab:system_baseline_comparison} compares the end-to-end quality--cost tradeoff across baselines. Relative to RAW logging, \sysname reduces storage from 195.4 to 43.3~MB/day ($4.51\times$) and average power from 5.29 to 0.68~mW ($7.78\times$), while achieving 0.825 mIoU and 34.192~dB PSNR. At nearly the same system cost, JPEC-Q30 requires 42.0~MB/day and 0.66~mW, but its quality drops to 0.514 mIoU and 27.29~dB PSNR. This large quality gap at comparable cost shows that aggressively compressing complete frames does not preserve the information most useful for downstream coral monitoring.

SeaScan further reduces storage to 9.2~MB/day and power to 0.63~mW, but operates at a much sparser 187-s system interval. However, it only consumes 0.6s on sensing and mostly spends time transmitting data, but SeaScan is not suitable for coral monitoring for its poor task performance; in comparison, \sysname observes the scene every 10~s while improving mIoU from 0.322 to 0.825 and PSNR from 28.23 to 34.192~dB. NeuriCam targets continuous 15-FPS video and consequently incurs substantially higher storage and power, making it poorly matched to long-term battery-powered monitoring. MCUCoder similarly provides neither lower system cost nor comparable task quality. Overall, \sysname achieves a stronger balance between observation frequency, downstream quality, and long-term resource cost.

\subsection{Seven-Day Coral-Tank Deployment}
\label{subsec:tank_deployment}

To validate the complete system beyond offline replay, we deploy the node continuously for seven days in a coral aquaculture tank 2~m deep $\times$ 1.5~m wide, as shown in \fig~\ref{fig:computing}(d). The recirculating reef ecosystem contains resident fish and dozens of soft and stony coral colonies, while water flow and illumination variation introduce natural motion and transient occlusions.
The node remains at a fixed viewpoint and follows the deployed 10-s monochrome and 180-s RGB schedule, producing 60,480 monochrome observations and 3,360 RGB anchors over seven days, or approximately 1.368~GB of source visual data before state-based reduction. From the recovered log, we report capture completeness, stored data volume, timestamp continuity, unexpected resets, corrupted records, microSD failures, and clock drift.

\head{Measured deployment energy} We measure battery consumption using an oscilloscope across a series sense resistor, convert the measured voltage drop to current, and integrate the current over time to obtain charge consumption in mAh. The seven-day steady-state operation consumes approximately 30.6~mAh, while setup consumes an additional 1.2~mAh (for we need time to first power on the system and then put it into a waterproof capsule).

\head{Deployment task performance} \fig~\ref{fig:case} shows representative reconstruction and segmentation results from the seven-day deployment. Across the deployment data, the system achieves 0.925 mean mIoU, 0.872 foreground IoU, 29.66~dB PSNR, and 0.991 SSIM. Segmentation is strong because each fixed view contains a dominant coral colony, simplifying the foreground compared with the multi-colony training scenes. Reconstruction preserves satisfied colony shape and local texture, although PSNR is lower than offline due to continuous water motion causing large geometric fluctuations even when the coral remains unchanged.

\vspace{-2mm}
\section{Discussion}
\label{sec:discussion}

\head{Data retrieval}
The current system stores data locally and is intended to be recovered after months of deployment, either by divers or underwater robots. Although retrieval still incurs cost, it replaces repeated site visits for routine data collection with only deployment and recovery operations.

\head{Multi-year deployment} 
While our prototype demonstrates a 160-day lifetime using a modest 1000~mAh battery, \sysname can practically support multi-year operations. First, outfitting the system with a larger, standard underwater battery canister (\eg, 10,000~mAh) would proportionally extend its lifespan to several years. 
Additionally, the battery life can also be further extended by, for example, reducing the baseline sampling frequency for slowly evolving reefs or aggressively duty-cycling the system into deep sleep during nighttime hours when passive illumination is insufficient.

\head{Low-cost online extension}
The current prototype operates fully offline to minimize underwater energy. Future versions could add a low-rate channel for heartbeat, battery and storage status, together with occasional event-triggered summaries. Full-resolution visual data would remain stored locally, preserving most of the lifetime benefit while providing basic online visibility into device and reef conditions.

\section{Related Work}
\label{sec:related}

\head{Conventional Coral Monitoring Platforms} (1) Diver surveys provide accurate colony-scale observations, but are labor-intensive and difficult to conduct frequently~\cite{nocerino2020coral,towle2022national}. (2) Robotic and fixed platforms reduce human effort: underwater robots expand survey coverage, while stationary systems such as SRMS and Coral City Camera enable repeated site-level observations~\cite{girdhar2023curee,maurya2023augmented,coralcitycamera}. (3) Remote sensing scales monitoring to reef and regional extents, including coral-recruitment habitat mapping, but water-column effects, spatial resolution, and revisit frequency constrain colony-scale, event-level monitoring~\cite{hedley2016remote,radford2024remote}. These limitations motivate persistent in-situ sensing that combines fine spatial detail with high temporal continuity.

\head{Ultra-Low-Power Underwater Imaging} CoralCam enables low-cost underwater time-lapse monitoring, demonstrating month-long deployment with twice-daily image capture~\cite{greene2020coralcam}. Battery-free underwater imaging eliminates batteries through acoustic energy harvesting and low-power color sensing with acoustic backscatter communication~\cite{afzal2022battery}. SeaScan further reduces underwater imaging cost using filtered monochrome sensors and backscatter, requiring as little as 23.6~mJ per color image~\cite{naeem2024seascan}. These systems optimize sparse capture or communication efficiency, whereas our work combines frequent low-power observations with sparse high-quality sensing and persistent on-device state for long-term ecological monitoring.

\head{MCU-Constrained Visual Computing and Storage} 
MobileNet and MCUNetV2 reduce embedded inference and memory cost through efficient operators and patch-based execution~\cite{howard2017mobilenets,lin2021mcunetv2}. Prior work also studies storage-energy reduction through batched writes and power management on embedded loggers~\cite{bradley2020optimising,rabault2026openwavelogger}, while MCUCoder brings learned compression to resource-constrained devices~\cite{hojjat2025mcucoder}. These works optimize computation, memory, compression, or storage, whereas long-term camera nodes must coordinate them under shared SRAM and energy budgets.

\head{Dual-Stream and Temporally Sparse Video Processing} NeuriCam~\cite{veluri2023neuricam} couples an always-on low-power grayscale stream with sparse high-resolution color keyframes, while SlowFast~\cite{feichtenhofer2019slowfast} separates slow semantic and fast motion pathways. DeltaCNN~\cite{parger2022deltacnn} exploits temporal coherence through sparse feature updates, and SkipNet~\cite{wang2018skipnet} conditionally skips redundant computation. These works inspire asymmetric sensing, temporal reuse, and conditional execution, but primarily target conventional video streams. Long-term duty-cycled underwater monitoring additionally requires explicit control of state staleness and refresh to prevent drift over extended deployments.

\head{Coral-Reef Visual Analytics and Task-Driven Restoration} Recent coral vision research has advanced toward dense and temporally consistent analysis. CoralSCOP provides coral-specific dense segmentation, CoralVOS extends it to video segmentation, and CoralSCOP-LAT supports automatic and semi-automatic labeling and analysis~\cite{zheng2024coralscop,zheng2025coralvos,wong2025coralscop}. Coral imagery also supports ecological indicators including cover, recruitment habitat, bleaching, disease, mortality, and other health changes~\cite{towle2022national,radford2024remote,kopecky2023quantifying}. Task-driven restoration similarly optimizes recovered observations for downstream recognition, detection, or segmentation rather than pixel-level fidelity alone~\cite{kim2024beyond,chen2025unirestore,kim2025exploiting}, while recent work selectively refines task-relevant latent tokens~\cite{lee2026tasktok}. Our framework follows this principle by prioritizing spatial and temporal evidence most useful for coral-monitoring tasks.

Existing work largely treats sensing, energy, computation, and coral analysis separately; our framework jointly designs them for persistent, task-aware coral monitoring.

\section{Conclusion}
\label{sec:conclusion}

We presented \sysname, an ultra-low-power coral monitoring system that exploits the temporal and spatial redundancy of fixed-view reef observations to avoid processing and storing every observation independently. By combining dual-rate sensing, persistent visual history, adaptive updates, and shore-side recovery, \sysname preserves frequent and high-fidelity coral reef observations at low underwater cost. Compared with baselines, it reduces storage by $4.51\times$ and average power by $7.78\times$, extending 1000mAh-battery lifetime to 159.5~days while maintaining strong coral-monitoring task performance, with 0.825 mIoU and 34.192~dB RGB PSNR.
A seven-day deployment in a real underwater coral ecosystem further demonstrates the practical deployability and reliable operation of \sysname under real-world conditions. 
We envision that \sysname underpins ultra-low-power, low-cost coral monitoring for ubiquitous and scalable deployment, 
enabling continuous visual evidence for studying long-term reef change, recovery, and conservation outcomes for sustainable ocean protection.


\balance
\bibliographystyle{unsrt}
\bibliography{refs}

\end{document}